\documentclass[12pt]{article}
\input{amssymb.sty}
\usepackage{graphicx}

\def\thefootnote{\fnsymbol{footnote}}
\catcode`\@=11
\def\numberbysection{\@addtoreset{equation}{section}
         \def\theequation{\thesection.\arabic{equation}}}
\numberbysection

\def\rrangle{\rangle\!\rangle}
\def\llangle{\langle\!\langle}

\newcommand {\ket}[1]{ \mbox{$\left|#1\right>$} }
\newcommand {\bra}[1]{ \mbox{$\left<#1\right|$} }
\newcommand {\ketI}[1]{ \mbox{$\left|#1\right\rrangle$} }
\newcommand {\braI}[1]{ \mbox{$\left\llangle#1\right|$} }
\def\beq{\begin{equation}}
\def\eeq{\end{equation}}
\def\bea{\begin{eqnarray}}
\def\eea{\end{eqnarray}}
\def\bdis{\begin{displaymath}}
\def\edis{\end{displaymath}}

\def\Z{{\Bbb Z}}

\def\E{{\cal E}}
\def\H{{\cal H}}
\def\R{{\cal R}}
\def\A{{\cal A}}
\def\D{{\cal D}}
\def\E{{\cal E}}

\renewcommand{\rho}{\varrho}
\begin{document}

\pagenumbering{alph}
\setcounter{page}{0}

\rightline{UCL--IPT--02--01}

\vskip 3cm
{\LARGE \centerline{Non periodic Ishibashi states :}
\centerline{the su(2) and su(3) affine theories }}

\vskip 2.5cm

\centerline{\large Philippe Ruelle\footnote{Chercheur
Qualifi\'e FNRS} and Olivier Verhoeven\footnote{Collaborateur scientifique
FNRS}}

\vskip 2truecm
\centerline{Institut de Physique Th\'eorique}
\centerline{Universit\'e Catholique de Louvain}
\centerline{B--1348 \hskip 0.5truecm Louvain-La-Neuve, Belgium}

\vskip 2truecm
\begin{abstract}
\noindent
We consider the $su(2)$ and $su(3)$ affine theories on a cylinder, from the
point of view of their discrete internal symmetries. To this end, we adapt
the usual treatment of boundary conditions leading to the Cardy equation to
take the symmetry group into account. In this context, the role of the Ishibashi
states from all (non periodic) bulk sectors is emphasized. This formalism is
then applied to the $su(2)$ and $su(3)$ models, for which we determine the
action of the symmetry group on the boundary conditions, and we compute the
twisted partition functions. Most if not all data relevant to the symmetry
properties of a specific model are hidden in the graphs associated with its
partition function, and their subgraphs. A synoptic table is provided that
summarizes the many connections between the graphs and the symmetry data that
are to be expected in general.
\end{abstract}

\renewcommand{\thefootnote}{\arabic{footnote}}
\setcounter{footnote}{0}
\newpage
\pagenumbering{arabic}
\baselineskip=16pt


\section{Introduction}

It is by now well--known that the content of two--dimensional conformal field 
theories (CFT) is best exposed on surfaces with distinct topology, an idea that
goes back to Cardy \cite{cardy1,cardy2}. In the same way open and closed strings
are related to each other, the study of a CFT on a surface with boundaries (a
cylinder say) is deeply connected with its formulation on a torus. Bulk data
and boundary data are inextricably related and subjected to numerous cross
consistency constraints \cite{c_lew,lew,pss}. Aspects of these connections can
be most elegantly expressed in terms of graphs \cite{dFZ,bppz1}: on a cylinder,
their nodes label the consistent boundary conditions and their adjacency
matrices yield the partition functions, whereas on the torus, their spectra
code the diagonal part of modular invariants.

Partition functions are fundamental objects characterizing a specific CFT. On
the torus, they are modular invariant and yield the field content in the
periodic sector. On the cylinder, they similarly give the field content in the
presence of boundary conditions. Torus and cylinder partition functions are not
independent: the Cardy equation \cite{cardy3} relates the boundary states
and the boundary partition functions to the periodic Hilbert space in the bulk.

The presence of a discrete symmetry within a CFT can be probed by allowing non
trivial monodromies of the fields along non contractible loops \cite{zuber},
leading to twisted partition functions. On the torus, these partition
functions must transform covariantly under the modular group, thereby relating
the bulk Hilbert spaces of the various twisted sectors. On the cylinder, the
analogous statement is a generalized Cardy equation, which relates the boundary
Hilbert spaces, and the action of the symmetry group on them, with the bulk
Hilbert spaces in the twisted sectors.

For the affine $\widehat{su}(2)$ and $\widehat{su}(3)$ theories, the (discrete)
symmetry group of each model has been determined in \cite{deux} from an
analysis on the torus.  The object of this article is to extend the study to the
cylinder, following a method presented in \cite{r} for the Virasoro minimal
models. As on the torus, where most of the information about the symmetry group
and its action on the fields was encoded in the graphs, we find here that the
same relationship holds on the cylinder. 

Therefore, instead of giving the detailed, case by case results of our
analysis, it is more appealing to say how they can be extracted from basic data
of the graph. To this end we provide in Table 2 a chart summarizing our
observations on the way the symmetry data can be read off from the graph
data. This table is a central result of this paper.

Before going to the cylinder, we start in Section 2 by briefly recalling the
highlights  of the analysis on the torus. A number of observations, made in the
Virasoro minimal models and in the $su(2)$ and $su(3)$ affine models and which
turn out to be crucial for the rest of the analysis, are explained (and
included in the synoptic table). 

Section 3 is devoted to a general discussion of the boundary conditions and of
the cylinder partition functions when the theory has a symmetry group. The main
result here is that a boundary condition invariant under a symmetry
subgroup can be expanded on the Ishibashi states taken from the torus sectors
twisted by that subgroup. It directly leads to a sequence of Cardy equations.

Relying in an essential way on the observations recalled in Section 2,
the fourth section and an appendix make explicit the action of the symmetry
group on the boundary conditions for the affine models considered here. 

For the same models, Section 5 solves the Cardy equations for the boundary
conditions which have a non trivial isotropy group, thereby determining the way
this group acts on the corresponding boundary Hilbert spaces. We show that the
relevant information can be extracted from subgraphs, made of the nodes left
fixed by the isotropy group.


\section{Symmetries in the bulk and graphs}

When a conformal field theory has a symmetry group $G$, twisted boundary
conditions around non--contractible loops may be considered \cite{zuber}. They
can be obtained from periodic boundary conditions by letting a group element
act before closing the loop, or from the insertion of suitable disorder fields.
The corresponding boundary conditions are labeled by group elements (there may
be others \cite{pz}). 

On a torus, there are two independent periods, chosen as 1 and $\tau$. Choosing
respectively $g$ and $e$ (i.e. periodic) boundary conditions along the two
periods yields the partition function of the $g$--twisted sector in the form
\beq
Z_{g,e} (\tau) = {\rm Tr}_{\H_g} [T_g]\, = \sum_{(j,j') \, \in \,\H_g}
M_{j,j'}^{(g)} \; \chi^*_j \; \chi_{j'},
\eeq
where $T_g=q^{L_0-{c \over 24}}\,\bar q^{\bar L_0-{c \over 24}}$ is the (finite
distance) transfer matrix with monodromy $g$. The twisted Hilbert space has been
decomposed as $\H_g = \oplus _{(j,j')} M_{j,j'}^{(g)}\, \R_j
\otimes \R_{j'}$ in terms of inequivalent representations of the symmetry
algebra, with (integer) multiplicities $M_{j,j'}^{(g)}$. 

A group element $f$ maps $\H_g$ to the isomorphic space $\H_{fgf^{-1}} \cong
\H_g$ since the conjugation by
$f$ changes the monodromy of the transfer matrix:
\beq
Z_{g,e} = {\rm Tr}_{\H_g} [f^{-1}f\,T_g \, f^{-1}f] = 
{\rm Tr}_{\H_{fgf^{-1}}} [T_{fgf^{-1}}] = Z_{fgf^{-1},e}.
\eeq

It follows that the centralizer of $g$ has an action into $\H_g$. The symmetry
means that the chiral algebra is left invariant by $G$, so that $G$ can only
rotate a whole representation to an equivalent one. Thus one finds 
\beq 
Z_{g,g'} (\tau) = {\rm Tr}_{\H_g} [T_g \, g']\, =\sum_{(j,j') \, \in \,\H_g}
\lambda_{j,j'}(g;g') \; \chi^*_j
\; \chi_{j'}, \qquad gg'=g'g,
\label{Zgg}
\eeq
with $\lambda_{j,j'}(g;g')$ the character of $g'$ acting on the
$M_{j,j'}^{(g)}$ degenerate representations in the sector $g$. This action is
unitary, and can be diagonalized, implying that all $\lambda_{j,j'}(g;g')$ can
be written as a sum of roots of unity (charges), of order equal to the order of
$g'$. 

The twisted partition functions should not be intrinsically sensitive to
a modular change of the modulus. Since a modular transformation mixes the two
periods, the invariance holds provided the boundary conditions are accordingly
changed. This results in the following identities
\beq
Z_{g,g'}(\tau) = Z_{g,gg'}(\tau+1) = Z_{g'^{-1},g}(-{\textstyle {1 \over
\tau}}).
\label{mod}
\eeq

The presence of a symmetry group can be asserted by finding a consistent set of
functions of the form (\ref{Zgg}) that transform covariantly under
the modular group as in (\ref{mod}). This set of functions contains in
particular the modular invariant partition function $Z_{e,e}$. For the affine
models based on $su(2)$ and $su(3)$, this analysis has been completed in
\cite{deux}. 

For the $su(2)$ affine models, classified by a list of ADE partition functions,
the symmetry group $G$ is the group of automorphisms of the Dynkin diagram A, D
or E, associated with the modular invariant, except for the $A_{n-1}$ models,
$n$ odd (that is, the diagonal theories for $\widehat{su}(2)_k$, $k$ odd),
which have no symmetry\footnote{In those cases, there is a symmetry
$\Z_2$ but it is realized projectively \cite{deux}.}. For the $su(3)$ models,
the symmetry groups are listed in Table 1 for all modular invariant partition
functions. 

\begin{table}
\renewcommand{\arraystretch}{1.6}
\hspace{-0.6cm}
\begin{center}
\mbox{\footnotesize
\begin{tabular}[t]{|c||c|c|c|c|c|c|c|c|c|c|}
\hline
MIPF & $A_n$ & $A^*_{4,2 m +1}$ & $A^*_{2 m \ge 6}$ & $D^{}_6=D_6^*$ &
$D^{}_9=D_9^*$ & $D_{3m\geq 12}$ & $D_{3 m \pm 1}$ & $D^*_{9 \neq n\ge 7}$ &
$E_8$ &
$E^*_8$
\\
\hline
\hline
$G$ & $\Z_3$ & $\Z_3$ & $\Z_3$ & $A_4$, $A'_4$ & $\Z_3$, $\Z_3'$ & $\Z_3$ &
$\Z_3$ &
$\Z_3$ & 
$\Z_3$ & $\Z_3$ \\
\hline
${\rm Aut}\,\Gamma^f$ & $\Z_3$ & $-$ & $\Z_2$ & $S_4$, $A_4 \times
\Z_2$ & $S_3$, $\Z_3$ & $S_3$ & $-$ & $\Z_3$ & $\Z_6$ & $\Z_2$ \\
\hline
\end{tabular}}
\end{center}
\begin{center}
\mbox{\footnotesize
\begin{tabular}[t]{|c||c|c|c|c|c|c|}
\hline
cont'd & $E_{12}$ & $E^{MS}_{12}$ & $E^{*MS}_{12}$ &
$E_{24}$\\
\hline
\hline
& $\Z_3$ & $-$ & $-$ & $-$ \\
\hline
& $S_3$, $S_3$, $S_3 \times \Z_3$ & $-$ & $\Z_2 \times \Z_2$ & $\Z_2$
\\
\hline
\end{tabular}}
\end{center}
\caption{Groups of symmetry pertaining to the $su(3)$ affine
models. The top line refers to the model (a modular invariant partition
function), in a notation borrowed from \cite{bppz2}. The second line mentions 
the symmetry group $G$ of the corresponding field theory. Two modular
invariants ($D_6$ and $D_9$) allow two different realizations of the same
symmetry group ($A_4$ and $\Z_3$ resp.). The third line gives the automorphism
group(s) of the generating graph(s) associated with it. }
\end{table}

Beyond the mere knowledge of the group, the partition functions must be given.
We will not reproduce the list here, which can be found in \cite{deux}, but we
simply recall the relations they bear with graphs. The observations we made
regarding these relations are summarized on the left half of Table 2, and will
play a essential role on the cylinder. 

\begin{table}[tpb]
\leavevmode
\begin{center}
\hspace{-1cm}
\mbox{\includegraphics[width=16cm]{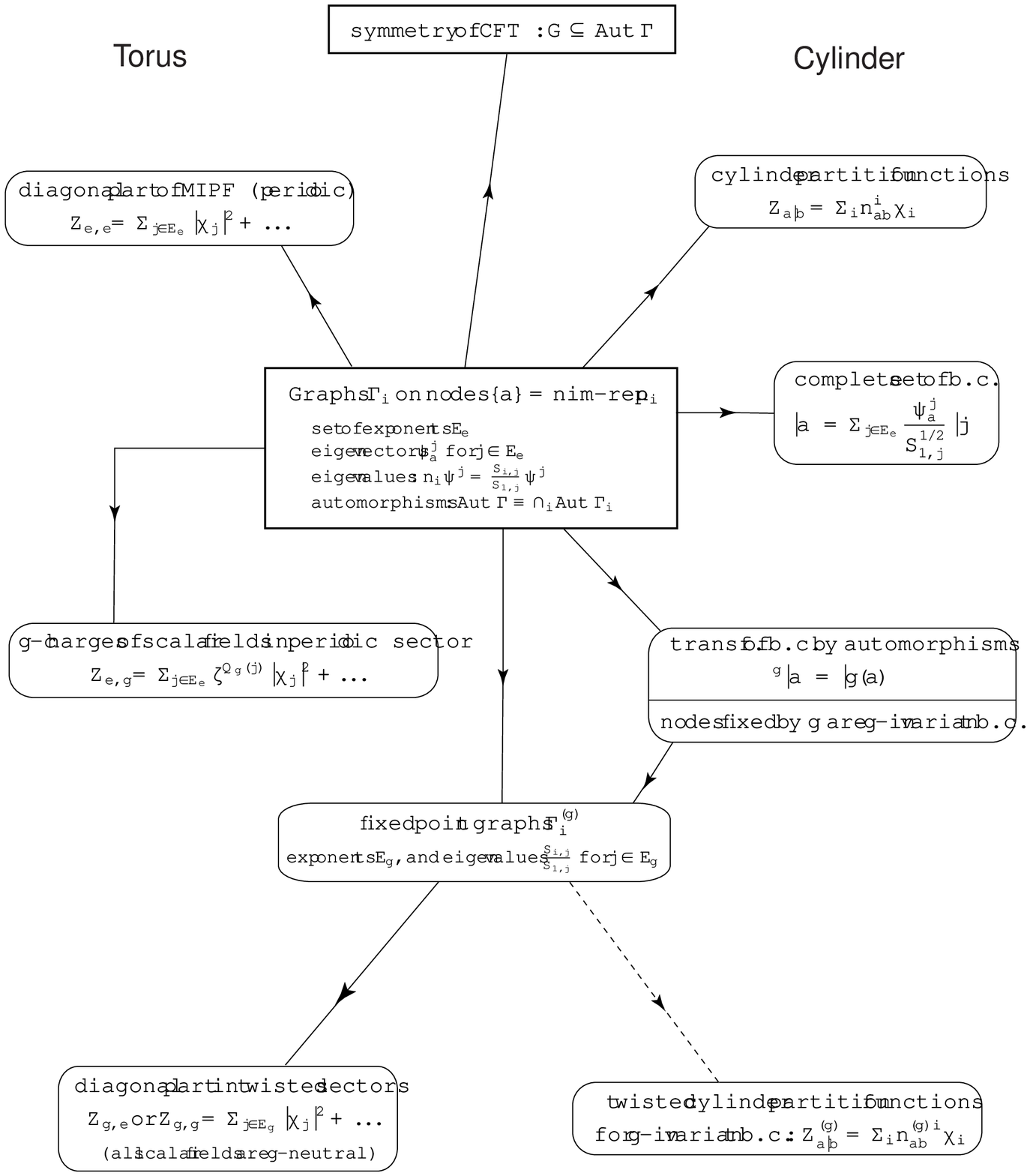}}
\end{center}
\caption{Synoptic table summarizing the links that can be expected between the
graphs and the various field theoretic data pertaining to an internal symmetry,
as suggested by our analysis of Virasoro minimal models, and $su(2)$ and
$su(3)$ affines models. All links are generically valid, although they are not
systematic, in the sense that specific prescriptions are sometimes needed. In
this respect, the arrow going from a fixed point graph to the twisted cylinder
partition functions requires specific prescriptions (to produce the phased
adjacency matrix $n^{(g)\,f}$ mentioned in the text), and for that reason, has
been dashed.}
\end{table}

Graphs were originally associated with a given modular invariant \cite{dFZ},
more precisely with its diagonal terms:
\beq
Z_{e,e}(\tau) = \sum_{(j,j) \in \H_e} \, \lambda_{j,j}(e;e) \, |\chi_j(\tau)|^2
+  \hbox{non--diagonal}.
\eeq
One first defines the set of exponents $\E_e = \{j \;:\; (j,j) \in \H_e\}$, 
where $j$ is taken in $\E_e$ with multiplicity $\lambda_{j,j}(e;e)$. Thus the
cardinal of $\E_e$ is the number of diagonal terms in $Z_{e,e}$. To this set
$\E_e$ one then associates a collection of graphs $\{\Gamma^i\}$ such that
the eigenvalues of the adjacency matrix $n^i$ of $\Gamma^i$ form the set
$\{{S_{i,j} \over S_{1,j}}\}$ for $j$ in $\E_e$, with $S$ the matrix
representing the modular transformation $\tau \to -{1 \over \tau}$ on the
characters. The number of graphs is equal to the dimension of the chiral fusion
ring, each graph living on a set of
$|\E_e|$ nodes, denoted $\{a\}$. From their spectrum, the matrices $n^i$ form a
non--negative integer representation of the fusion ring (a ``nimrep''). Among
them one can choose a set of generators, and the corresponding graphs, which we
will call the generating graphs. For the $su(2)$ and $su(3)$ affine models, the
fusion ring has one generator $n^f$, so that a single graph $\Gamma^f$ needs be
given for each modular invariant (the index $f$ can be chosen to be a
fundamental representation).

For $su(2)$, there is a unique generating graph associated with a modular
invariant of A, D or E type, and it is precisely the corresponding Dynkin
diagram \cite{bppz2}. For $su(3)$, one generating graph has been found for each
modular invariant, but in three cases, several (isospectral) graphs
are known \cite{dFZ}: the two self--conjugate invariants $D_6^{}=D_6^*$ and
$D_9^{}=D_9^*$ can each be associated two different generating graphs, usually
denoted by $\D_6^{},\D_6^*$ and $\D_9^{},\D_9^*$, while for the exceptional
invariant $E_{12}$, three isospectral graphs $\E_{12}^{(1)}, \E_{12}^{(2)},
\E_{12}^{(3)}$ exist. The resulting list of graphs, likely to be complete but
not proved to be so, can be found in \cite{bppz2}.

It turns out that the symmetry group $G$ in the affine models is very close to
the automorphism group of the generating\footnote{When there is more than one
generator, one should consider instead the symmetries $\sigma$ of the nimrep
$n^i_{ab}=n^i_{\sigma(a),\sigma(b)}$, or equivalently the intersection $\cap_i
$ Aut$\,\Gamma^i$.} graph $\Gamma^f$. The two groups are equal in all $su(2)$
models except those of even rank $A$ type. For $su(3)$, two types of graphs are
to be distinguished. The fold graphs\footnote{They are graphs obtained by
quotienting other graphs by automorphisms that act freely. The graphs
mentioned in the text are the fold graphs of respectively $\D^*_n$, $\A_{3m\pm
1}$ and $\E_8$ by a $\Z_3$ group.} associated to the theories $A^*_n$,
$D_{3m\pm 1}$ and $E^*_8$ have an automorphism group (often trivial) with no
direct relation to $G$. For all other graphs, $G$ is either equal to or is a
subgroup of Aut $\Gamma^f$. (See Table 1 for a comparison of $G$ and
Aut$\,\Gamma^f$ in all cases.)

The relevance of the graph is strengthened by the following two observations
\cite{r,deux}. 

\begin{enumerate}
\item
For the non--fold graphs, $G$ is a subgroup of the graph automorphisms. As such
it acts on an eigenbasis $\psi^j$ of the adjacency matrices $n^i$. It leaves
invariant the various eigensubspaces, labeled by the exponents of $\E_e$ which
are different, and so it acts by a block diagonal representation $R(g) =
\oplus_{j\;:\; (j,j) \in \H_e} \; R_j(g)$, where $R_j$ has dimension
$M^{(e)}_{j,j}=\lambda_{j,j}(e;e)$ (the multiplicity of $j$ in $\E_e$). Being
the symmetry group of the conformal theory, $G$ also acts on the periodic
Hilbert space, in particular on the diagonal representations of
$\H_e$. This action is again block diagonal since $G$ can only mix degenerate
$\R_j \otimes \R_j$, also labeled by the distinct exponents. Thus $G$ acts on
$(\H_e)_{\rm diag}$ through a reduced representation $R'(g) = \oplus_{j\;:\;
(j,j) \in \H_e} \; R'_j(g)$ of the same form as $R(g)$.\\
Our first observation is that $R'(g)$ coincides\footnote{This is a minor point
but in fact, one of the two is going to be an anti-representation
(the transpose of a representation), see Eq. (\ref{transf}). On the other hand,
$R'$ is known only through its character (from the partition functions).} with
$R(g)$ (up to the choice of basis, so they are equivalent). Therefore the
numbers
$\lambda_{j,j}(e;g)$ that appear in front of the diagonal terms of $Z_{e,g}$
are exactly the characters of the automorphism $g$ acting on the eigenspaces of
the matrices (graphs) $n^i$
\beq
Z_{e,g} = \sum_{j\,:\, (j,j) \in \H_e} \; \lambda_{j,j}(e;g) \, |\chi_j|^2 +
\ldots 
\qquad
\Leftrightarrow \qquad \lambda_{j,j}(e;g) = {\rm ch}\,R_j(g).
\label{eigen}
\eeq
Therefore the $G$--charges of the scalar fields in the periodic sector are
entirely dictated by the graph (the charges are noted $\zeta^{Q_g(j)}$ in 
Table 2; they are the eigenvalues of the representation $R_j(g)$ or $R'_j(g)$ of
which $\lambda_{j,j}(e;g)$ is the character).

For the fold graphs of $su(3)$, it turns out that all scalar fields in $Z_{e,e}$
are invariant under the action of $G$ (all
$\lambda_{j,j}(e;g)=\lambda_{j,j}(e;e)$ for all $g$). Then the above observation
still holds if we view the whole of $G$ as having no action at all on the
graphs. The same view will be suggested when going to the cylinder.

\item
It is very natural to define sets of twisted exponents $\E_g = \{j \;:\;
(j,j) \in \H_g\}$, where each $j$ is again taken with the multiplicity of
$(j,j)$ in $\H_g$, and for each $g$, to look for matrices $n^{(g)\,i}$ whose
spectrum is given by the set $\{{S_{i,j} \over S_{1,j}}\}$ for $j$ in $\E_g$.
Clearly, the  $n^{(g)\,i}$ again satisfy the fusion algebra and are
polynomially expressible in terms of a few generators, in our case, just one,
namely $n^{(g)\,f}$. It no longer is a positive integer matrix in general, but
when its entries are sums of roots of unity, as will soon be the case, one
can see it as the adjacency matrix of a phased graph.\\ 
A simple inspection of the torus twisted partition functions reveals the
striking fact that $|\E_g|$ is equal to the number of graph nodes that are fixed
by $g$
\beq
|\E_g| = \#\{a\;:\; g(a) = a\}.
\label{fixed}
\eeq
Moreover, in many cases, the restriction $\left. \Gamma^f\right|_g$ of the
generating graph $\Gamma^f$ to the nodes fixed by $g$ has the right spectrum
$\{{S_{i,j} \over S_{1,j}}\}_{j \in \E_g}$.  This is quite non--trivial in view
of the wild spectral changes that a restriction usually causes. In those happy
cases, one could set $n^{(g)\,f} = \left. n^f\right|_g$ and generate all
$n^{(g)\,i}$ from it\footnote{In these cases, $n^{(g)\,f}$ is still positive
integer--valued but the other matrices $n^{(g)\,i}$ will not be positive (albeit
integral).}. Finally, it was noted that for $g \neq e$, all diagonal terms of
$Z_{g,e}$ are invariant under the centralizer of $g$, except for the $D_6^{}$
and $D_6^*$ invariants of $\widehat{su}(3)_3$ (which are the only cases where
the centralizer of $g$ is not generated by $g$). 

\end{enumerate}

The second observation, which extends the first one to the non periodic
sectors, is very suggestive, but as such, remains vague. In a sense it is the
purpose of this article to give them a full and precise meaning on the
cylinder. As we shall see, these observations will have very strong
consequences on the determination of the cylinder partition functions. 


\section{Boundary conditions}

The previous section spelt out the consequences of an internal symmetry for the
formulation on a torus, and its connection to graphs. The question we ask now is
what this analysis becomes on a cylinder. 

\begin{figure}[htb]
\begin{center}
\mbox{\includegraphics[width=4.7cm]{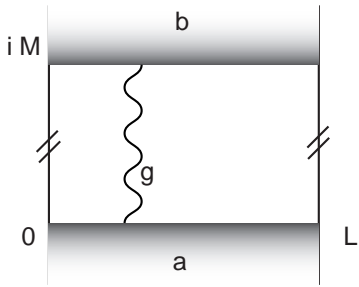}}
\end{center}
\caption{Geometric setting for the calculation of the cylinder twisted
partition function $Z_{b|a}^{(g)}$.}
\end{figure}

We view the cylinder of perimeter $L$ and length $M$ as a rectangle, with
vertical sides identified. There are two boundaries, the two horizontal sides,
on which boundary conditions $a$ and $b$ are prescribed. We will require that
they preserve the chiral algebra present, here an affine algebra. The way
boundary conditions can be handled is by now standard
\cite{cardy3,c_lew,lew,bppz2}; we will adapt the usual presentation in order to
include the symmetry. 

The symmetry group acts on all the states. In particular, its acts on the
boundary conditions $a \to {}^g a$, and on the states that can circulate the
non--trivial cycle. The latter may have a non trivial monodromy around that
loop, which, as before, can be implemented by the insertion of a disorder
line, running from one boundary to the other (see Figure 1).

As usual, there are two equivalent ways of computing the partition
function corresponding to boundary conditions $a,b$ and monodromy $g$.
If constant time slices are the closed horizontal circles, the proper
Hilbert space to consider is $\H_g$, encountered above on the torus (and thus
known). On the boundaries, the boundary conditions prescribe
boundary states $\ket{a}$ and $\ket{b}$. Then the partition function is
\beq 
Z^{(g)}_{b|a} = \bra{b} e^{- M H_g} \ket{a} = \bra{b} e^{\frac{-2 \pi
M}{L}(L_0 + \bar{L}_0 -\frac{c}{12})} \ket{a}.
\label{Zabg1}
\eeq

If instead the time is chosen to run along the horizontal coordinate, the
space corresponds to the segment of length $M$, with boundary conditions
$a$ and $b$ at the two ends. The Hilbert space is now some $\H_{b|a}$
(independent of $g$). The time translation is periodic with period
$L$, resulting in the partition function
\beq
Z^{(g)}_{b|a} = {\rm Tr}^{}_{\H_{b|a}} [g\, e^{- L H_{b|a}}] = {\rm
Tr}^{}_{\H_{b|a}} [g\,e^{-\frac{\pi L}{M} (L_0-\frac{c}{24})}].
\label{Zabg2}
\eeq
Various constraints on $a,b$ and $\H_{b|a}$ arise from the equality of these
two forms (moreover $g$ may be varied independently over the symmetry group
$G$).

The first form (\ref{Zabg1}) shows that both boundary states $a$ and $b$ should
have a projection in the space $\H_g$. If the projection of either one
vanishes, so does the partition function $Z_{b|a}^{(g)}$. In the
second form (\ref{Zabg2}), the states of $\H_{b|a}$ are transported around the
perimeter, acted on by $g$, and projected on themselves. The action of $g$
turns the states of $\H_{b|a}$ into states of $\H_{{}^g b|{}^g a}$, so that the
trace gives zero if the two boundary conditions $a$ and $b$ are not invariant
under $g$ (the form (\ref{Zabg2}) would not be well--defined and would depend
on the position of the disorder line). 

Combining the two observations, one learns that boundary states belong
to the Hilbert spaces $\H_g$ for those $g$ which leave them invariant :
\beq
\ket{{}^g a} = \ket{a}  \qquad \Longleftrightarrow \qquad \ket{a} \in \H_g.
\eeq

The boundary states preserving the chiral algebra must satisfy the
linear equations $[L_n - \bar L_{-n}] \ket{a} = 0$ for all $n \in \Z$, for
conformal invariance, and similar equations with the affine generators $J^c_n$
for the affine invariance. They form a vector space, where a convenient basis is
provided by the so--called Ishibashi states $\ketI{j}$. Each such state
$\ketI{j}$ satisfies the linear equations and belongs entirely to the diagonal
representation $\R_j \otimes \R_j$, and vice--versa each diagonal
representation $\R_j \otimes \R_j$ contains exactly one Ishibashi state
$\ketI{j}$. These solutions form a complete set of solutions (over $\Bbb C$). 

The boundary conditions are required to belong to various Hilbert
spaces $\H_g$, according to the above discussion. In each $\H_g$ (they are all
known), a basis is provided by the Ishibashi states belonging to that
particular space, and we denote them by $\ketI{j}_g$. They are 
labeled by the diagonal representations occurring in $\H_g$, that is, by the
set called $\E_g$ in the previous section (the diagonal terms of $Z_{g,e}$).
The way these states are constructed shows in addition that
\beq
{}_g\braI{j} q^{{1 \over 2}(L_0+\bar L_0-{c \over 12})} \ketI{j'}_{g'} = 
\chi_j(q) \, \delta_{j,j'} \, \delta_{g,g'}.
\eeq

Consequently a boundary state invariant under a set of group elements $g$
possesses many alternative writings, since for every such $g$, it can be written
as a linear combination of Ishibashi states from the $g$--twisted sector,
conventionally written with a factor of $S_{1,j}$,
\beq
\ket{a} = \sum_{j \, \in \, \E_g} {\psi^{(g)\,j}_a \over \sqrt{S_{1,j}}} \,
\ketI{j}_g,
\label{a}
\eeq
for some complex coefficients $\psi^{(g)\,j}_a$ to be determined.

The previous two equations allow to compute the partition function in the first
form (\ref{Zabg1}):
\beq
Z_{b|a}^{(g)} = \sum_{j \,\in\, \E_g} \; {\psi^{(g)\,j}_a \, \psi^{(g)\,j*}_b
\over S_{1,j}} \,  \chi_j(\tilde q)\,, \qquad \tilde q = e^{-{4\pi M \over L}}.
\label{tilde}
\eeq

In order to compute the second form of it, one decomposes the space $\H_{b|a}$
into inequivalent representations of (a single copy of) the chiral algebra,
$\H_{b|a} = \oplus_i \, n^i_{ba} \R_i$, with $n^i_{ba}$ integers. If $g$ leaves
$a$ and $b$ invariant, it also commutes with the Hamiltonian $H_{b|a}$, and
acts unitarily in $\H_{b|a}$ by rotating the equivalent representations, as
before. Denoting by $n^{(g)\,i}_{ba}$ the character of the representation
$L^i_{b|a}$ through which $g$ acts on the $n^i_{ba}$ equivalent representations
($n^i_{ba}=n^{(e)\,i}_{ba}$), one simply has from (\ref{Zabg2})
\beq
Z^{(g)}_{b|a} = \sum_i \; n^{(g)\,i}_{ba} \, \chi_i(q)\,, \qquad q=e^{-{\pi L
\over M}}.
\label{q}
\eeq

Writing $\tilde q = e^{2i\pi \tilde\tau}$ and $q = e^{2i\pi\tau}$,
the numbers $\tilde\tau$ and $\tau$ are related by a modular transformation.
Transforming the partition function (\ref{tilde}) in the basis of characters
$\chi_i(q)$, one finds\footnote{As is well--known, some care is needed. The
identification of the coefficients of the characters relies on their linear
independence, which, for current algebras, means using the unspecialized
characters. This can be done by including sources (related to the Cartan
subalgebra generators) in the stress--energy tensor, see the Appendix A in
\cite{bppz2}. The explicit calculation in particular shows that the
characters in (\ref{tilde}) are related to those in (\ref{q}) by an $S^{-1} =
S^*$ transformation. This results in a complex conjugation in the Cardy
equation.} the Cardy equation, generalized to account for the symmetries,
\beq
(n^{(g)\,i}_{ab}\big)^* = \sum_{j \,\in\, \E_g} \; \psi^{(g)\,j}_a \,
{S_{i,j} \, \over S_{1,j}} \, \psi^{(g)\,j*}_b\,, \qquad \qquad {\rm for\ }{}^g
a=a,\, {}^g b=b.
\label{cardy}
\eeq
The matrices satisfy $n^{(g)\,i^\star} = (n^{(g)\,i})^\dagger$ for pairs
$i,i^\star$ of conjugate fields. The other relation $n^{(g)\,i} =
(n^{(g^\star)\,i})^*$ follows if there is an involution $g \to g^\star$, $a
\to a^\star$ such that $\E_g^\star = \E_{g^\star}$ and $(\psi^{(g)\,j}_a)^* =
\psi^{(g^\star)\,j^\star}_{a^\star}$. The involution is of course trivial for
$su(2)$. One checks that it also is trivial for the $su(3)$, except for the
models $A^*_n$, $D_n^{}$ and $E_8^*$ where $g^\star=g^{-1}$ ($g$ of order 3).
Whenever $g^\star \neq g$, the set of numbers $\{{S_{i,j} \, \over S_{1,j}}\}_{j
\in \E_g}$ is not closed under complex conjugation.

By definition, the boundary conditions are those combinations of Ishibashi
states such that all the numbers $n^i_{ab}$ are non--negative integers. As
positive integral combinations of boundary conditions are again boundary
conditions, it is convenient to find a set of extremal, {\it pure} ones (in
analogy with the extremal equilibrium states), of which the others are positive
superpositions. A usual though somewhat strong working hypothesis is to assume
the existence of an orthonormal and complete set of pure boundary conditions,
satisfying
\beq
\sum_{j \in \E_e} \; \psi^{(e)\,j\,*}_a \, \psi^{(e)\,j}_b = \delta_{a,b}\,,
\qquad \sum_{a} \; \psi^{(e)\,j\,*}_a \, \psi^{(e)\,j'}_a = \delta_{j,j'}\,.
\label{compl}
\eeq
It is by no means obvious that complete boundary conditions exist in all
sensible rational (even unitary) conformal theories. Indeed infinitely many
examples of affine (WZW) theories are known where an orthonormal and complete
set does not exist \cite{gannon}; they have however not been proved (or
disproved) to be sensible theories in the first place. In a number of cases
however, including the minimal Virasoro models, and the $su(2)$ and
$su(3)$ models examined here, an orthonormal and complete set does exist.

The conditions (\ref{compl}) imply $n^1_{ab}=\delta_{a,b}$ from Eq.
(\ref{cardy}), that is, the identity field occurs once in the space $\H_{a|a}$,
and is absent from $\H_{b|a}$ for $a \neq b$. They also fix the
number of pure boundary conditions to be $|\E_e|$. 

More importantly, they show that the coefficients $\psi^{(e)\,j}_a$ are the
entries of a unitary matrix. Thus from Eq. (\ref{cardy}) for $g=e$, all matrices
$n^i$ are diagonalized by $\psi^{(e)}$ and have their spectrum given by the set
$\{{S_{i,j} \, \over S_{1,j}}\}_{j \in \E_e}$. Thus they form the nimreps
discussed in the previous section, and can be seen as the adjacency matrices of
graphs, the $\psi^{(e)\,j}_a$ being their common eigenvectors. The nodes of the
graphs are thus seen to be labels for pure boundary conditions. 

The completeness condition leads to a reformulation of the Cardy equation
for $g=e$ as the problem of finding graphs with prescribed spectra \cite{bppz1}.
Solving that problem yields an explicit determination of the boundary conditions
themselves and all the partition functions $Z^{}_{b|a} \equiv Z_{b|a}^{(e)}$. It
remains to determine the functions $Z_{b|a}^{(g)}$ for the other $g$'s.

Before doing that, let us note that the partition functions have the expected
symmetries $Z^{}_{b|a} = Z^{}_{{}^g b| {}^g a}$, valid for all boundary
conditions and all $g$ in $G$. This is easily seen from a double insertion of
$g$ and $g^{-1}$ on two lines parallel to the perimeter of the cylinder, with
say the $g$--line close to the $a$ boundary, and the $g^{-1}$--line close to the
$b$ boundary. The two insertions amount to no effect at all, but each line can
be pulled over to the closest boundary, transforming the boundary condition. 


\section{Symmetry of boundary conditions}

For solving the Cardy equation for the other $g$'s, one first needs to determine
the way a symmetry group element $g$ acts on the (pure) boundary conditions.
For doing that, the observations of Section 2 are essential (we emphasize
that we have no proof that these hold in all generality). 

The previous section gives us the boundary conditions as graph nodes and
as elements of $\H_e$,  i.e. as linear combinations of Ishibashi states
$\ketI{j}_e$ from the periodic sector. The diagonal terms of the torus
partition function $Z_{e,g}$ say how these states transform under the symmetry
group $G$, namely through the
$R'$ representation (see Section 2).  The observation we made was that this
action precisely coincides with the action of $g$, seen as an automorphism of
the graphs, on the eigenvectors of the adjacency matrices $n^i$. Introducing a
degeneracy index $\alpha_j$ taking values between 1 and the multiplicity of
the exponent $j$, one easily deduces
\beq
{}^g\ket{a} =  \sum_{j \in \E_e} \frac{\psi^{(e)\,j}_a}{\sqrt{S_{1,j}}}  \;
{}^g \ketI{j}_e  = \sum_{j,\alpha_j,\alpha'_j}
\frac{\psi^{(e)\,(j,\alpha_j)}_{a}}{\sqrt{S_{1,j}}}
\; [R'_j(g)]_{\alpha_j,\alpha'_j}\; \ketI{j,\alpha'_j}_e = \sum_{j \in
\E_e} \frac{\psi^{(e)\,j}_{g(a)}}{\sqrt{S_{1,j}}} \;\ketI{j}_e  = \ket{g(a)}.
\label{transf}
\eeq
Thus the action of a symmetry $g$ on the boundary conditions $\{a\}$ is
simply the action of the corresponding automorphism on the graph nodes. In 
particular, the $g$--invariant boundary conditions correspond to the nodes that
are left fixed by the automorphism associated to $g$. From Eq. (\ref{fixed}),
their number is equal to $|\E_g|$. We noted in Section 2 that for the $su(3)$
models associated to fold graphs ($A^*_n$, $D_{3m \pm 1}$ and $E^*_8$), all
representations $R'_j$ are trivial. It follows that all boundary conditions are
invariant in those cases, and we can set $g(a)=a$ for all $a$.

All $g$--invariant boundary conditions are linear combinations of Ishibashi
states taken from $\H_g$. The unknown coefficients $\psi^{(g)\,j}_a$
are constrained by the Cardy equation
\beq
(n^{(g)\,i}_{ab}\big)^* = \sum_{j \,\in\, \E_g} \; \psi^{(g)\,j}_a \, {S_{i,j}
\, \over S_{1,j}} \, \psi^{(g)\,j*}_b\,.
\label{cardy2}
\eeq
As in the usual case $g=e$, a solution to this equation yields at the same
time the coefficients $\psi^{(g)}$ and the matrices $n^{(g)\,i}$. There is
however an additional global condition on these matrices, which expresses the
fact that $n^{(g)\,i}_{ab}$ result from the action of a group element $g$. 

Define $G_a \subset G$ to be the isotropy group of the boundary condition $a$.
Then for $G_{a,b} = G_a \cap G_b$ the isotropy group of the pair $a$ and $b$,
the numbers $n^{(g)\,i}_{ab}$ for fixed $i,a$ and $b$, must be the character of
a representation of $G_{a,b}$. As discussed in the previous section, $G_{a,b}$
acts in the Hilbert space $\H_{a|b} = \oplus_i \, n^i_{ab} \R_i$, by a block
diagonal representation $L_{a|b} = \oplus_i \, L_{a|b}^i$. Then, one must have
($n^i_{ab} = n^{(e)\,i}_{ab}$) 
\beq
\forall i \;: \quad n^{(g)\,i}_{ab} = {\rm ch}\,L_{a|b}^i(g)\,, \qquad \forall g
\in G_{a,b}.
\label{char}
\eeq 

For the affine theories based on $su(2)$ and $su(3)$, the data pertaining to
$g=e$ are assumed to be known: the matrices $n^i$ are adjacency matrices of
graphs, satisfying the fusion algebra, and $\psi^{(e)}$ is their
diagonalizing matrix. Our task is to find the data for the other $g$'s
satisfying the above conditions, Eqs. (\ref{cardy2}) and (\ref{char}). This is
done in the following section, and relies heavily on the same graphs that
give the data for $g=e$.

Before that, we make a last comment regarding the transformation
of the boundary conditions. That certain boundary conditions can be
viewed as elements of different Hilbert spaces $\H_g$ makes sense, as a boundary
condition may be compatible with various monodromies. A concrete
example is the free boundary condition in the Ising model, which is not more
periodic than antiperiodic.

However their transformation under $G$, in Eq. (\ref{transf}), has been obtained
from their expression as elements of $\H_e$. It showed which of them are
invariant under which $g$'s, and in turn, this enabled us to write them as
elements of other spaces $\H_g$. Consistency requires to make sure that these
alternative writings are compatible with the transformations used to define
them. Using once again the observations of Section 2, it is not difficult, but
instructive, to see that they are indeed consistent (see however a peculiarity
for the $D_6^{}$ and $D_6^*$ models of
$su(3)$). The arguments are presented in an appendix.


\section{Solutions to the twisted Cardy equation}

The number of boundary conditions invariant under a given group element
$g$ is equal to $|\E_g|$, the number of diagonal terms in the twisted partition
functions $Z_{g,e}$. An immediate consequence is that all matrices $\psi^{(g)}$
are square. 

We have $n^1_{ab} = \delta_{a,b}$ from the completeness condition. This implies
$n^{(g)\,1}_{ab} = \delta_{a,b}$ for all $g$--invariant boundary conditions
$a,b$, because it must be diagonal (compatibility with $n^1_{ab}$), with
real positive entries (from Eq. (\ref{cardy2})). It means that
$\psi^{(g)}$ is unitary, and this in turn implies that all
$(n^{(g)\,i})^*$ are diagonalizable, have a spectrum equal to the set of ratios 
$\{\frac{S_{i,j}}{S_{1,j}}\}_{j\in \E_g}$, and therefore form a representation
of the fusion algebra. It also means that the $g$--invariant pure boundary
conditions (nodes fixed by $g$) form a complete set of $g$--invariant boundary
conditions.

This puts the twisted data $n^{(g)\,i}$ on an equal footing as the untwisted
$n^i$, with of course several important differences. First, the $n^{(g)\,i}$
form representations of the fusion ring of (much) lower dimension than the
$n^i$. Second, the matrices $n^{(g)\,i}$ for fixed $g$, do not yield a nimrep,
but instead a $\Z(\zeta_N)$--valued representation of the fusion algebra, with
$\zeta_N=e^{2i\pi \over N}$ and $N$ the order of $g$. One may interpret them as
adjacency matrices of phased graphs, in which links are assigned phases.

In addition, all $n^{(g)\,i}$ must be compatible with $n^i$, in the sense of
Eq. (\ref{char}). It implies, for every $g \in G_{a,b}$ (the subgroup 
leaving the boundary conditions $a$ and $b$ invariant), that
\beq
(n^{(g)\,i}_{ab})^* = n^i_{ab} \;\;\bmod (1-\zeta_N).
\eeq
This compatibility condition can be put in yet another but more suggestive
form: for fixed $g$, the matrix $(n^{(g)\,i})^*$, with indices in the set of
$g$--invariant boundary conditions, is equal to the restriction of $n^i$ to the
nodes fixed by $g$, up to phases ($N$--th roots of unity).

This firmly suggests that solutions $(n^{(g)\,i})^*$ to the Cardy equation can
be obtained from those subgraphs, by going through the following steps:
\begin{itemize}
\item[(i)] consider the subgraphs $\Gamma^{(g)\,i}$ corresponding to the
nodes fixed by $g$ and their adjacency matrices; 
\item[(ii)] weigh the adjacency matrices by appropriate phases so as to
reproduce the correct spectra $\{\frac{S_{i,j}}{S_{1,j}}\}_{j\in \E_g}$ (that
is, attach every link of the subgraphs a phase rather than the number 1,
thus obtaining a phased graph);
\item[(iii)] then check the compatibility of the resulting matrices with the
$n^i$. 
\end{itemize}

This procedure has been used to obtain the solutions $n^{(g)\,i}$ presented
below in terms of the generator $n^{(g)\,f}$ of the fusion algebra
($n^{(g)\,f^\star} = (n^{(g)\,f})^\dagger$ for $su(3)$). It turns out that in
$su(2)$, $n^{(g)\,f}$ is the plain restriction of $n^f$ to the nodes fixed by
$g$, i.e. no weight on the links. On the other hand, non trivial weights are
required in $su(3)$ in order to achieve the desired spectra. 

This procedure however does not guarantee the uniqueness of the solutions it
produces. The generating matrices $n^{(g)\,f}$ are primarily constrained by
their spectrum, so that a unitary redefinition $n^{(g)\,f} \to U\,n^{(g)\,f}
\,U^\dagger$ is possible (it amounts to $\psi^{(g)} \to U\psi^{(g)}$), provided
the entries of $U$ are combinations of $n$--th roots of unity, so that the new
matrices be still $\Z(\zeta_N)$--valued.

The other constraints on $n^{(g)\,f}$ are the group properties with respect to
$g$ : their entries must be characters, Eq. (\ref{char}). A general unitary
rotation replaces $n^{(g)\,f}_{ab}$ by a linear combination
$\sum_{a',b'}\,U^{}_{aa'}\,U^*_{bb'} \, n^{(g)\,f}_{a'b'}$ taken over all
sectors $a',b'$ invariant under $g$. It is very implausible that these new
combinations will have the character property. Moreover they mix characters
from different sectors of boundary conditions, which is very unnatural (the
various boundary conditions being mixed have also different isotropy groups in
general).

Altough we have no formal proof of that, the only reasonable
unitary transformations are by diagonal matrices, which transform
$n^{(g)\,f}_{ab}$ into $\varphi_a^{} \varphi_b^* \, n^{(g)\,f}_{ab}$, for some
roots of unity $\varphi_a$ (which suitably depend on $g$ so as to keep the
character property). At least, we could verify in all models but the $D_n$
series of $su(3)$ that no other transformation is allowed. 

We will not dwell on this phase freedom. It has been discussed in the 
Virasoro minimal models in \cite{r}, where the Perron--Frobenius theorem was
used to determine these phases, relying on the assumption that the
non--degeneracy of the groundstate of the transfer matrix carried over in the
continuum limit. The affine models however are not known to be the scaling
limits of critical lattice models. Moreover, in $su(3)$ models, the groundstate
turns out to be degenerate for many boundary conditions. Consequently there is
no natural way to fix the freedom in the phases. (We note nonetheless that all
solutions for $su(2)$ given below, satisfy the Perron--Frobenius criterion.)


\subsection{The su(2) twisted partition functions}

The $su(2)$ models are particularly simple since the subgraphs of nodes fixed
by a symmetry are very small, containing one or two nodes. Only the $D_n$
theories have a large, unbounded number of fixed points.

The theories with a symmetry group are the series of $A_{n-1}$ for $n$ even, all
$D_{{n \over 2}+1}$ for $n \geq 6$ and $E_6$, which all have a $\Z_2$ symmetry,
except the theory $D_4$ which has an $S_3$ symmetry. The subgraphs fixed by the
$\Z_2$ form, respectively, an $\A_1$ graph, an $\A_{{n \over 2}-1}$ graph and an
$\A_2$ graph. The $D_4$ theory has in addition a single node fixed by the
$\Z_3$ (and by the whole of $S_3$).

In all cases, one may verify that the adjacency matrix of the subgraph
$\Gamma^{(g)\,f}$ of nodes fixed by $g$ provides a solution $n^{(g)\,f}$ to the
Cardy equation and satisfies the group property (\ref{char}). The other matrices
$n^{(g)\,i} = U_{i-1}(n^{(g)\,f})$ are generated from
$n^{(g)\,f}$ via the Tchebychev polynomials. They are in general different
from the restriction of the $n^i$ to the fixed subgraphs. 

The simplest way to see that they have the right spectrum is to observe that all
fixed subgraphs are $\A_{n'-1}$ graphs for a level $n'={n \over d}$ which
divides the level $n$ of the original theory, and that the set of twisted
exponents is exactly equal to $\E_g = d \, \E_e(A_{n'-1})$ (see \cite{deux}
for the sets $\E_g$). 


\subsection{The su(3) twisted partition functions}

We quickly review the results for the $su(3)$ models. We again distinguish two
classes of theories. One class contains the models $A_n^*$, $D_{3m \pm 1}$ and
$E_8^*$, which are those associated with a fold graph (unlike the $su(2)$
notations, the subscript refers here to the height $k+3$). The other class to
be discussed includes the models $A_{3m}$, $D_{3m}$, $D_6^*$ and $E_{12}$. All
other models either have no symmetry at all, or no invariant boundary
conditions.

The main difference between the two classes is that the models in the first
one have all their boundary conditions invariant under the symmetry. They are
also more easily handled.

For $i=(a,b)$ the integrable highest weights, we denote the automorphisms
(simple currents) of $\widehat{su}(3)_k$ as $\mu(a,b)=(n-a-b,a)$, with respect
to which the $S$ modular matrix transforms as $S_{\mu(i),j} = \omega^{t(j)} \,
S_{i,j}$, where $t(j)$ is the triality of $j$ and $\omega=e^{2i\pi/3}$. 

In what follows, we will take the sets of exponents $\E_g$ as given by the
twisted partition functions computed in \cite{deux}.

\subsubsection{First class : the fold graphs}

All models in this class have $\Z_3$ as symmetry group, and thus three sets of
exponents $\E_e$, $\E_g$ and $\E_{g^2}$. In all cases, these three sets have
the same cardinality, and are explicitely given as $\E_{g^2} = \E_g^\star$ and
$\E_g = \mu^r(\E_e)$, with $r=1$ or 2 depending on the models. For instance,
for the $D_{3m \pm 1}$ theories, we find $\E_e = \{i \;:\; t(i)=0\}$, $\E_g =
\{i= \;:\; t(i)=\mp 1\}$ and $\E_{g^2} = \{i= \;:\; t(i)=\pm 1\}$, so that $\E_g
= \mu^2(\E_e)$. The other cases have $\E_g = \mu(\E_e)$.

A direct consequence is that the spectra of $(n^{(g)\,i})^*$  and $n^i$ are
related by
\beq
\left\{ \frac{S_{i,j}}{S_{1,j}} \right\}_{j \; \in \; \E_g} = 
\left\{ \frac{S_{i,\mu^r(j)}}{S_{1,\mu^r (j)}} \right\}_{j \; \in \; \E_e} = 
\omega^{rt(i)} \left\{ \frac{S_{i,j}}{S_{1,j}}\right\}_{j \; \in \; \E_e},
\eeq
so that $n^{(g)\,i} = \omega^{-rt(i)} n^i$ is clearly a solution to the Cardy
equation, with $\psi^{(g)}=\psi^{(e)}$. It is also manifestly compatible with
$n^i$. 

We note that, in the graphs associated to all these theories, some of the links
are unoriented, which means that the corresponding sector of boundary
conditions has a degenerate groundstate ($n^f_{ab} = n^{f^\star}_{ab}$).

\subsubsection{Second class : the colourable graphs}

One finds in this class the $A_{3m}$, $D_{3m \geq 9}$ and $E_{12}$ theories, all
with a $\Z_3$ symmetry, and also the $D_6^{}$ and $D_6^*$ theories, both with
an $A_4$ symmetry. 

\medskip
{\it (i)} The series $A_{3m}$ is trivial: the node $(m,m)=({n \over 3},{n \over
3})$ is the only one that is invariant under $\Z_3$. Since $S_{f,(m,m)}=0$,
Cardy's equation says that $n^{(g)\,f} = n^{(g^2)\,f} = 0$ is the only solution
(which coincides with the restriction of $n^f$ to the fixed node). 

\medskip
{\it (ii)} The $E_{12}$ theory has three nodes invariant under the $\Z_3$, and
the spectrum of $n^{(g)\,f}$ is found to be $\{1,\omega,\omega^2\}$, the three
third roots of unity. Three distinct graphs $\E_{12}^{(i)}$ have been associated
to the $E_{12}$ modular invariant. Together, they yield two different solutions
$n^{(g)\,f}$ to the Cardy equation:
\bea
&& {\rm graph}\;\E_{12}^{(1)} \;: \qquad n^f\Big|_{\rm fixed} = \pmatrix{0&2&0
\cr 0&0&2 \cr 1&0&0} \quad \longrightarrow \quad n^{(g)\,f} = \pmatrix{0&-1&0
\cr 0&0&-1 \cr 1&0&0},\\
&& {\rm graphs}\;\E_{12}^{(2,3)} \;: \qquad n^f\Big|_{\rm fixed} =
\pmatrix{0&1&0 \cr 0&0&1 \cr 1&0&0} \quad \longrightarrow \quad n^{(g)\,f} =
\pmatrix{0&1&0 \cr 0&0&1 \cr 1&0&0}.
\eea
For the first one, the matrix $n^{(g)\,f}$ corresponds to assigning the phases
$\omega$ and $\omega^2$ to the two double links.

\medskip
{\it (iii)} In the case of $D_6^{}$ and $D_6^*$, the group elements in $A_4$
have order 2 or order 3. The theory $D^*_6$ has no invariant boundary
condition under order 3 elements, while $D^{}_6$ has, but is treated below,
along with the whole series $D^{}_{3m}$. For order 2 elements, the two
theories $D_6^{}$ and $D_6^*$ behave in a similar way, and can be treated
simultaneously.

The group $A_4$ has three elements of order 2, all conjugate to each other.
The discussion is the same for any of them (only the nodes that are invariant
change in $D^*_6$). Picking one particular $g$ of order 2, we find two
invariant boundary conditions (in the $D_6$ model, they are invariant under
the full $A_4$). The set of twisted exponents is in both cases
$\E_g=\{(2,2),(2,2)\}$, yielding a spectrum of $n^{(g)\,f}$ equal to 
$\{{S_{f,j} \over S_{1,j}}\}_{j \in \E_g} = \{0,0\}$. Cardy's equation implies
$n^{(g)\,f}=0$ identically, which is the restriction of $n^f$ in the case of
$D_6^*$. For $D_6$, the restriction of $n^f$ is equal to $\Big({{0 \;\;2} \atop
{0 \;\;0}}\Big)$, and $n^{(g)\,f}$ simply corresponds to assigning the double
link the phases $+1$ and $-1$.

\medskip
{\it (iv)} It remains to discuss the series $D_{3m}$, for which 
\bea
&& \E_e = \{j \;:\; t(j)=0 \} \cup \{(m,m),(m,m)\},\\
&& \E_g= \{j \;:\; t(j)=2 \} \,, \qquad \E_{g^2}= \{j \;:\; t(j)=1 \}
=\E_g^\star.
\eea

Let us recall that the graph $\D_{3m}$ is obtained from $\A_{3m}$ by an
orbifold procedure. Decomposing all weights $i$ of the alc\^ove as $B_{3m} =
T_0 \cup T_1 \cup T_2 \cup \{(m,m)\}$, with $T_k = \mu(T_{k-1})$, the
orbifolding triplicates the $\mu$--fixed point $(m,m)$, and establishes the
identifications $i \sim \mu(i)$ in the three sets $T_k$. The geometric
symmetry $\Z_3$ of the graph $\A_{3m}$ is broken by these identifications, but
is restored in $\D_{3m}$ because the three nodes coming from the triplication
of $(m,m)$ are symmetrical (the graph has a symmetry $S_3$). The
resulting graph $\D_{3m}$ (its adjacency matrix) has a spectrum that is the
zero triality subset of that of $\A_{3m}$, modulo the fact that $(m,m)$ occurs
three times. 

All nodes of $\D_{3m}$ except the triplicated ones are fixed points under the
$\Z_3$ symmetry, and in number equal to any $|T_k|$. It is easy to see that a
`phased' orbifold procedure can be defined that instead selects the triality one
or two subset of the spectrum of $\A_{3m}$, and leads to the phased
graphs with the spectrum corresponding to the exponents $\E_g$ or $\E_{g^2}$.  

According to the above decomposition $B_{3m} = T_0 \cup T_1 \cup T_2 \cup
\{(m,m)\}$, the adjacency matrix of the graph $\A_{3m}$ can be written
\beq \label{An}
\A_{3m}
 = \left( \begin{array}{cccc} 
A_0 & A_1 & A_2 & \alpha\\
A_2 & A_0 & A_1 & \alpha\\
A_1 & A_2 & A_0 & \alpha\\
\beta & \beta & \beta & 0\\
\end{array} \right)\,,
\eeq
where $\alpha$ and $\beta$ are respectively column and row vectors. Since this
graph encodes the fusion by the fundamental $f$, its eigenvalues are
$\{{S_{f,j} \over S_{1,j}}\}_{j \in B_n}$. With respect to the same
decomposition, its eigenvectors $\psi^{(j)}_i = S_{j,i}$ can be written as
$(v^{(j)};\omega^{t(j)}\,v^{(j)};\omega^{2t(j)}\,v^{(j)};S_{j,(m,m)})$ with
$v^{(j)} = (S_{j,i})_{i \in T_0}$. The eigenequation implies in particular
\beq
[A_0 + \omega^{t(j)} \, A_1 + \omega^{2t(j)} \, A_2] \, v^{(j)} + \alpha \,
S_{j,(m,m)} = {S_{f,j} \over S_{1,j}} \, v^{(j)}.
\eeq
The restriction to those $j$ of triality 2 shows that the matrix $A_0 + \omega^2
A_1 + \omega A_2$ has a spectrum equal to $\{{S_{f,j} \over S_{1,j}}\}_{j \in
\E_g}$. Similarly one finds that $A_0 + \omega A_1 + \omega^2 A_2$ has a
spectrum equal to $\{{S_{f,j} \over S_{1,j}}\}_{j \in \E_{g^2}}$. Thus the two
matrices yield solutions to the twisted Cardy equation
\beq
n^{(g)\,f} = A_0 + \omega A_1 + \omega^2 A_2\,, \qquad 
n^{(g^2)\,f} = (n^{(g)\,f})^* = A_0 + \omega^2 A_1 + \omega A_2.
\eeq

In terms of phased graphs, the solutions $n^{(g)\,f}$ and $n^{(g^2)\,f}$
involve a few phased links ($2m-3$), namely those which connect the
sets $T_0$ and $T_1$, as illustrated in Figure 2.

\begin{figure}[htb]
\leavevmode
\begin{center}
\mbox{\includegraphics[width=12cm]{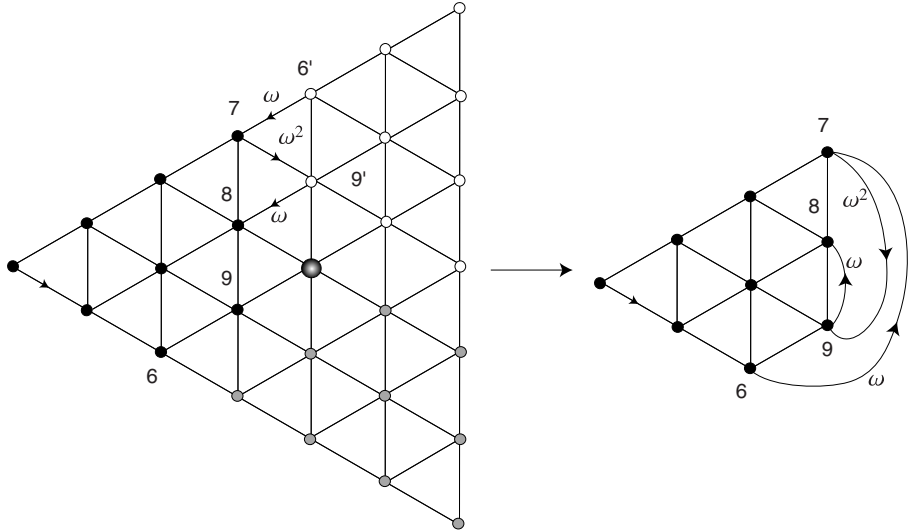}}
\end{center}
\caption{Shown on the right is the phased graph $(n^{(g)\,f})^*$ for $\D_9$,
as resulting from a twisted orbifold of the graph $\A_9$. The three colours
differentiate the three domains $T_k$ related by $\mu$ automorphisms, under
which the central node is fixed. All the links are oriented.}
\end{figure}


\vskip 1truecm
\appendix
\section{Boundary conditions and twisted Ishibashi states}

A pure boundary condition $a$ can be expanded in the basis of Ishibashi
states from the periodic sector. We used this expansion to determine their
transformation properties under the symmetry group $G$. We have then argued
that a boundary condition that is invariant under a subgroup $G_a$ of
$G$ can be expanded in Ishibashi states from all sectors twisted by
elements of $G_a$:
\beq
\ket{a} = \sum_{j \, \in \, \E_g} {\psi^{(g)\,j}_a \over \sqrt{S_{1,j}}} \,
\ketI{j}_g, \qquad g \in G_a.
\label{twa}
\eeq
For consistency, all these expressions for $g \neq e$ must be compatible with
the transformations of $a$ under $G$. We show that it is indeed the case
(except in one instance). The arguments below apply equally well to the minimal
Virasoro models.

We fix $a$ and a $g$ in $G_a$, and we consider the action of an arbitrary
element $f$ of $G$ on the expression of $\ket{a}$ as element of $\H_g$, Eq.
(\ref{twa}). We distinguish the two cases according to whether $f$ is in the
centralizer
$C(g)$ of
$g$ or not. 

An $f \not\in C(g)$ transforms $\ketI{j}_g$ into an Ishibashi state of
$\H_{fgf^{-1}}$. Assuming no mixing (by a proper choice of basis), that is,
$f\ketI{j}_g = \ketI{j}_{fgf^{-1}}$, we obtain
\beq
^f\ket{a} = \sum_{j \, \in \, \E_g} {\psi^{(g)\,j}_a \over \sqrt{S_{1,j}}} \,
\ketI{j}_{fgf^{-1}}\,.
\label{exp1}
\eeq

We note that the boundary state $\ket{f(a)}$ has isotropy group $G_{f(a)} =
fG_af^{-1}$, and can therefore be written as
\beq
\ket{f(a)} = \sum_{j \, \in \, \E_{fgf^{-1}}} {\psi^{(fgf^{-1})\,j}_{f(a)}
\over \sqrt{S_{1,j}}} \, \ketI{j}_{fgf^{-1}}\,,
\label{twfa}
\eeq
for the same $g$. The Hilbert space $\H_{fgf^{-1}}$ is isomorphic to $\H_g$ so
that the set of exponents $\E_{fgf^{-1}}$ is equal to $\E_g$. Moreover the
coefficients $\psi^{(fgf^{-1})\,j}_{f(a)}$ are related, through the Cardy
equation, to the numbers $n^{(fgf^{-1})\,i}_{f(a),f(b)}$ specifying the (trace
of the) action in the space $\H_{f(a)|f(b)}$ of the isotropy group 
$G_{f(a)} \cap G_{f(b)}$. This action is clearly conjugate to the action of $G_a
\cap G_b$ in the Hilbert space $\H_{a|b} \cong \H_{f(a)|f(b)}$. From this it
follows that $n^{(fgf^{-1})\,i}_{f(a),f(b)} = n^{(g)\,i}_{a,b}$, and that the
matrices $\psi^{(fgf^{-1})\,i}$ and $\psi^{(g)\,i}$ are equal (may be chosen to
be equal). Thus $\ket{f(a)}$ can also be written as
\beq
\ket{f(a)} = \sum_{j \, \in \, \E_g} {\psi^{(g)\,j}_{a}
\over \sqrt{S_{1,j}}} \, \ketI{j}_{fgf^{-1}}\,.
\label{exp2}
\eeq
Comparing (\ref{exp1}) and (\ref{exp2}), one sees that $^f\ket{a}$ coincides
with one of the possible writings of $\ket{f(a)}$. (Note that $f$ may or may
not be in $G_a$. A typical example is the $D_4$ invariant of $su(2)$, having
$G=S_3$. The boundary condition corresponding to the middle node in the
Dynkin diagram is fully invariant under $S_3$: it can be written, in three
different ways, as an element of the sector twisted by an order two
element, and in two different ways, as an element of the sector twisted by an
order three group element. The $\Z_2$ expressions get permuted by the $\Z_3$
subgroup of $S_3$, and vice--versa.)

If $f$ is in the centralizer of $g$, it acts in $\H_g$ (hence on the
$\ketI{j}_g$) in a way that can be read off from the torus partition function
$Z_{g,f}$. It turns out, in all but two cases, that (i) the centralizer of $g$
is the subgroup generated by $g$ (hence $f$ is a power of $g$), and (ii) all
diagonal terms of $Z_{g,e}$ are invariant under $g$ hence under $f$ (i.e.
$\lambda_{j,j}(g,f)=1$ in the notations of Section 2). Together, these two
facts imply that $\ket{a}$, written as an element of
$\H_g$, is invariant under $f$. 

The two cases where points (i) and (ii) above do not hold are the $D^{}_6$ and
$D^*_6$ theories of $\widehat{su}(3)_3$, when $g$ is an element of order 2 in
$G=A_4$. The group $A_4$ has three elements of order 2, all conjugate, forming
with the identity a subgroup\footnote{The modular invariant itself can be viewed
as the reduction of the diagonal invariant of $\widehat{so}(8)_1$, in which
$\widehat{su}(3)_3$ is conformally embedded. The subgroup $\Z_2 \times \Z_2$
survives the embedding and corresponds to the group of simple currents of
$\widehat{so}(8)$. In terms of $\widehat{so}(8)$ primary fields however, the
twisted torus partition functions contain no diagonal term.}
$\Z_2 \times \Z_2 \equiv \{e,a,a',aa'\}$. On the torus, both models $D^{}_6$ and
$D^*_6$ carry the same action of this subgroup (they are distinguished by the
realization of the $\Z_3$ subgroup). 

The centralizer of any order 2 element is equal to $\Z_2 \times \Z_2$.
The partition functions in the sector twisted by $a$ say, read \cite{deux}
(those in the sectors $a'$ or $aa'$ are similar)
\beq
Z_{a,a^k a'^l} = [1+(-1)^l]\, |\chi_{(2,2)}|^2 + (-1)^k \, 
\Big\{\chi_{(2,2)}^* \, [\chi_{(1,1)}+\chi_{(4,1)} +\chi_{(1,4)}] 
+ (-1)^l \times {\rm c.c.} \Big\},
\eeq
for $k,l=0,1$. Concentrating on the two degenerate, diagonal terms,
one sees that they are invariant under $a$, but have opposite charges under
$a'$. 

This has the following consequence. From the graphs $\D_6^{}$ and $\D^*_6$, one
sees that in each case, there are two nodes invariant under $a$ (in $\D_6$, they
are actually invariant under the full $A_4$). The corresponding two boundary
states, call them 1 and 2, can thus be expressed as linear combination of the
two Ishibashi states of the sector twisted by $a$ ($S_{(1,1),(2,2)}={1 \over
2}$) :
\bea
& & \ket{1} = \sqrt{2} \, \Big[\psi^{(a)\,(2,2)}_1 \, \ketI{(2,2)}_a + 
\psi^{(a)\,\widetilde{(2,2)}}_1 \, |\widetilde{(2,2)}\rangle\!\rangle_a \Big],
\\ & & \ket{2} = \sqrt{2} \, \Big[\psi^{(a)\,(2,2)}_2 \, \ketI{(2,2)}_a + 
\psi^{(a)\,\widetilde{(2,2)}}_2 \, |\widetilde{(2,2)}\rangle\!\rangle_a \Big].
\eea

In $D_6^*$, the graph says that the two boundary states are interchanged under
$a'$ or $aa'$. Assuming without loss of generality that $\ketI{(2,2)}_a$ and
$|\widetilde{(2,2)}\rangle\!\rangle_a$ have respective charges $+1$ and $-1$
under $a'$ (both are $+1$ under $a$), the matrix of coefficients $\psi^{(a)} =
{1 \over \sqrt{2}} \Big({1 \;\;\;\; 1 \atop 1 \;\; -1}\Big)$ makes the
transformations $(a'\ {\rm or}\ aa')\ket{1}=\ket{2}$ and $(a'\ {\rm or}\
aa')\ket{2}=\ket{1}$ manifest, and is unitary. 

In $D_6^{}$ on the contrary, the two boundary states are invariant under $a'$
and $aa'$, so one would like to determine the coefficients to make these
invariances manifest. However there is no way to do it with a unitary matrix
$\psi^{(a)}$. The reason for this is unclear to us.


\end{document}